\documentclass[pra,superscriptaddress,twocolumn]{revtex4-1}
\usepackage{amssymb,amsmath,amsthm}
\usepackage{color}
\usepackage{bm}
\usepackage{dsfont}
\usepackage{hyperref}
\usepackage{graphicx}

\definecolor{darkblue}{rgb}{0,0,0.6}
\definecolor{darkred}{rgb}{0.5,0,0}

\hypersetup{
 pdfborder={0,0,0},
 colorlinks=true,
 linkcolor=darkblue,
 urlcolor=darkblue,
 citecolor=darkblue
}

\begin{document}

\title{On the connection between the theorems of Gleason and of Kochen
  and Specker}
\author{Karl-Peter Marzlin}
\affiliation{Department of Physics, St. Francis Xavier University,
  Antigonish, Nova Scotia, B2G 2W5, Canada}
\author{Taylor Landry}
\affiliation{Department of Physics, St. Francis Xavier University,
  Antigonish, Nova Scotia, B2G 2W5, Canada}

\begin{abstract}
We present an elementary proof of
a reduced version of Gleason's theorem and
the Kochen-Specker theorem to provide a novel perspective on the
relation between both theorems.
The proof is based on a set of linear equations for the values of a function $m$
on the unit sphere.
In the case of Gleason's theorem the
entire unit sphere needs to be considered, while
a finite set of points suffices to prove the Kochen-Specker theorem.
\end{abstract}

\maketitle

\section{Introduction}
Quantum theory is a spectacularly successful description of 
the dynamics of atoms and molecules and has been confirmed
in countless experiments. 90 years after de Broglie proposed
matter waves, quantum mechanics still fascinates us because
it is so profoundly different from classical mechanics 
and sometimes seems to defy common sense.
Two of the most famous theorems that pinpoint the differences between
classical and quantum theory are those of Gleason \cite{GleasonsTheorem}
and of Kochen and Specker \cite{KochenSpecker68}. In a nutshell,
Gleason proved that the probability $p(\psi) = |\langle \psi
|\sigma\rangle |^2 $ to find a system in state $|\psi \rangle $ 
when it has been prepared in state $|\sigma \rangle $ follows
from a small number of rather general assumptions.
Kochen and Specker showed that
it is impossible to assign a value to all observables 
simultaneously. This is in contrast to classical theories, where
observables always assume a specific value, even if we may not know
this value. The physical and philosophical implications of
both theorems have been described in many publications.
An overview can be found in Refs.~\cite{Redhead:Incompleteness,Hemmick2012}

It is well-known that both theorems are connected and that
the Kochen-Specker theorem may be considered as a corollary
of Gleason's theorem \cite{Granstrom2006}. However, their
proofs are of very different nature. 
The proof by Kochen and Specker can be reduced to showing that
it is impossible to color the unit sphere with two colours in a
particular way. Gleason's
proof, on the other hand, has been described as 
``famously difficult'' \cite{Bengtsson:GleasonKS}.
The theorem has since been proven
in different ways
\cite{PSP:2091976,hellmanGleason1993,billingeGleason1997,Richman1999287,RichmanGleason2000}
and has been extended to open quantum systems 
\cite{jmp/35/12/10.1063/1.530679,PhysRevLett.91.120403,CavesGleason2004,WstarGleason}
and to quantum information \cite{Edalat2004,Wallach0050028}.

If both theorems are closely connected, why is the result
of Gleason so much more difficult to obtain? 
The purpose of this paper is to
answer this question in a simple way
that is also accessible to undergraduate students.

\section{Gleason's Theorem}\label{sec:gleason}
We consider a variant of Gleason's 
theorem that has been discussed
by Gudder (corollary 5.17 of Ref.~\cite{GudderQM1979}).
\\[3mm]
{\bf Reduced Gleason Theorem}: Let ${\cal H}$ be a real separable Hilbert space of
dimension $\geq 3$ and ${\cal P}({\cal H})$ the lattice of projectors
(see App.~\ref{app:projectors})
on ${\cal H}$. Let $m$ be a map ${\cal P}({\cal H})\rightarrow
[0,1]$ which satisfies 
\begin{align} 
  m(\hat{\mathds{1}})&=1
\label{eq:assump1}\\
  m\left (\sum_i \hat{P}_i \right ) &=\sum_i m\left (\hat{P}_i \right ) 
   \; \text{for mutually orthogonal } \hat{P}_i.
\label{eq:assump2}\end{align} 
Furthermore, we assume that a rank-1 projector $\hat{P}_\sigma$ exists
such that $m(\hat{P}_\sigma)=1$. Then
$m(\hat{P})= \text{Tr}( \hat{P}\hat{P}_\sigma) $ 
for all $\hat{P} \in {\cal P}({\cal H})$.
\\[3mm]
We begin by discussing the main differences to Gleason's full
theorem.
First, we have chosen to consider a real Hilbert space because
it is suitable for our purpose.
Below we show that if the Kochen-Specker theorem
holds for a real Hilbert space, then it also holds for a complex
Hilbert space. To establish a connection between Gleason's theorem and
the real Kochen-Specker theorem, the
reduced form of Gleason's Theorem is sufficient. In addition,
a real Hilbert space is advantageous for pedagogical purposes.
We remark that a real Hilbert space can still capture many, though not
all, aspects of quantum mechanics.
For instance,  (nonlocal) violations of the Bell inequality 
\cite{Bell:Physics1964,RevModPhys.38.447}, which are often used
in quantum information to test entanglement 
\cite{Tittel2011:Nature09719}, can be obtained on a real Hilbert space
\cite{Redhead:Incompleteness}. Another example is quantum chemistry,
where the vast majority of calculations employ real
superpositions of real electron orbitals \cite{SzaboOstlund1996}.

A second difference to Gleason's full theorem is that we assume
the existence
of a  rank-1 projector $\hat{P}_\sigma$ for which
$m(\hat{P}_\sigma)=1$. In the literature such an $m$ is called an
atomic state. Gleason showed that such a function $m$ represents
the same information as the (pure) state $|\sigma \rangle  \in {\cal H}$ 
on which $\hat{P}_\sigma$ projects. Furthermore, Gleason proved the existence
of $|\sigma \rangle$, rather than assuming it, and thus showed that
the usual expression for quantum mechanical mean values,
$m(\hat{P})= \langle \sigma |\hat{P}|\sigma \rangle $, is unique
under the assumptions of his theorem. Gleason also considered
mixed states, but for the purpose of a comparison of the two
theorems we will concentrate our efforts on pure states.

In most applications, the map $m$ represents the probability
distribution for observables represented by projectors, and
$|\sigma \rangle $ describes the state in which the system
is prepared. Clearly, the probability to find the system in the state 
 $|\sigma \rangle $ in which it has been prepared must be unity, so
that  $m(\hat{P}_\sigma)=1$, where $\hat{P}_\sigma$ is the projector
on the subspace spanned by $|\sigma \rangle $. Also, the probability to find any state
at all must be 1, which is the statement of Eq.~(\ref{eq:assump1}).

Projectors that project on orthogonal subspaces are commuting and can therefore be
measured simultaneously. Eq.~(\ref{eq:assump2}) expresses the fact
that such measurements are statistically independent, so that the
respective probabilities can be added.

Gleason's theorem is an extremely powerful result. 
The axioms of quantum mechanics include the statement that
if a system is prepared in state $|\sigma \rangle $, then the
probability to find it in state $|\psi \rangle $ is given by
$p(\psi) = |\langle \psi |\sigma\rangle |^2 $. 
If $\hat{P}_\psi $ denotes the projector on vector $ |\psi \rangle $,
then this probability can also be expressed in the form
$p(\psi) = \langle \sigma| \hat{P}_\psi  |\sigma\rangle $. 
What Gleason achieved is to reduce the axiomatic framework 
of quantum theory: if we accept that the probability
to find the system in a state $|\psi \rangle $ is somehow related to 
$\hat{P}_\psi$, then his theorem
completely fixes $p(\psi)$.

There is one physical assumption behind Gleason's theorem that
is not obvious from its mathematical statement:
non-contextuality. To understand what this means, imagine we
try to measure whether the spin of an electron points in the positive
$z$-direction. Mathematically, this measurement can be described
by a projector $\hat{P}_z$. Physically, the Zeeman effect
implies that we should employ a magnetic field $\vec{ B}_z$ pointing in the 
$z$-direction for this experiment. On the other hand, if we measure
whether the spin points in the positive $x$-direction (projector $\hat{P}_x$),
a magnetic field $\vec{ B}_x$ pointing in the  $x$-direction would be needed.

The map $m(\hat{P})$ in Gleason's theorem is non-contextual in the
sense that it does not depend on how the measurement is performed:
we use the same map $m$ regardless of whether we consider
$\hat{P}_z$ or  $\hat{P}_x$. 
However, in quantum physics
measuring non-commuting observables requires a different
experimental setup, so that there is no compelling reason why $m$ should
be the same.
In a contextual theory, $m$ would depend both
on the projector and on all physical
parameters needed to perform the experiment. In our spin example,
a contextual theory would consider a map $m(\hat{P}, \vec{ B})$
rather than $m(\hat{P})$. Such a change would
ruin the proof of Gleason's theorem as presented below.

This point may seem a bit meticulous, but it has important consequences.
Much work has been devoted to the question whether quantum mechanics
can be interpreted in the terms of classical probability theories by introducing
``hidden variables'' (HV), i.e., parameters that may affect an experiment
but to which we have no access. 
If $m(\hat{P})$ represents the
probability to find the system in in the subspace associated with
$\hat{P}$, then Gleason's theorem 
can be used to show that non-contextual HV theories
cannot be in agreement with the results predicted
by quantum theory \cite{PhysRevA.69.022118,PhysRevA.89.032123}. 
However, it does not exclude 
contextual HV theories
\cite{RevModPhys.38.447,Redhead:Incompleteness,Hemmick2012}.
In the discussion of
  the Kochen-Specker theorem below, we will return to contextuality
  and provide a refined definition that is more amenable for 
 quantum theory.

\section{Proving the reduced Gleason theorem}\label{sec:GleasonProof}
The fundamental idea behind the proof is to find a set of
orthogonal vectors such that assumptions 
(\ref{eq:assump1}) and (\ref{eq:assump2}) can only be
fulfilled for a unique function $m(\hat{P})$. We will do this
in several steps: (A) show that working in a 3D space is sufficient,
(B) show that $m$ can only depend on the scalar product
$\langle \psi|\sigma \rangle $  between a vector
$|\psi \rangle $ and the prepared state $|\sigma \rangle $, and
(C) show that this function of the overlap must take the form
given in the theorem. Our proof starts in a similar way as that of
Gudder \cite{GudderQM1979} and is inspired by some of the techniques
used in Refs.~\cite{Granstrom2006,MorettiGleasonSphericalHarmonics}.

\subsection{Reduction to 3D and some Lemmas} 
Our goal is to derive the value of $m(\hat{P}_\chi)$ for a specific
vector $|\chi \rangle \in {\cal H}$. If $|\chi \rangle $ is proportional to
$|\sigma \rangle $, we have $m(\hat{P}_\chi)=1$. In all other cases,
 $|\chi \rangle $ and $|\sigma \rangle $ span a two-dimensional
subspace of ${\cal H}$, for which we can use a basis consisting of
the two vectors $ |\sigma \rangle$ and $ |\sigma_\perp \rangle $.
For technical reasons we will need a third dimension
\footnote{There is a counterexample for Gleason's theorem and the
  Kochen-Specker theorem in the two-dimensional case
  \cite{Redhead:Incompleteness}, where $m(\hat{P})$ can be
considered as a function $m(\varphi)$ of the angle $\varphi$ on the
unit circle. The choice $m(\varphi) = 0$ for $\varphi \in [0, \frac{
  \pi}{2}) \cup [\pi, \frac{ 3\pi}{2})$ and $m(\varphi) = 1$ elsewhere 
then fulfills assumptions (\ref{eq:assump1}) and (\ref{eq:assump2})
but contradicts the statement of both theorems.
} and 
therefore introduce
a third orthonormal normalized vector $|\sigma'_\perp \rangle $
that is perpendicular to both $|\sigma \rangle $ and $|\sigma_\perp
\rangle $. Any
normalized vector $|\psi \rangle $ in this 3D subspace of ${\cal H}$ can then be written as
\begin{align} 
  |\psi(\theta,\varphi) \rangle  = \cos\theta |\sigma \rangle + \sin\theta \cos\varphi
  |\sigma_\perp \rangle 
  +\sin\theta \sin\varphi |\sigma'_\perp \rangle .
\label{eq:psiGeneral}\end{align} 
Hence, $\cos\theta$ corresponds to the overlap $\langle \sigma | \psi \rangle $
between $|\psi(\theta,\varphi) \rangle$ and $|\sigma \rangle $.

\begin{figure}
\begin{center}
\includegraphics[width=6.5cm]{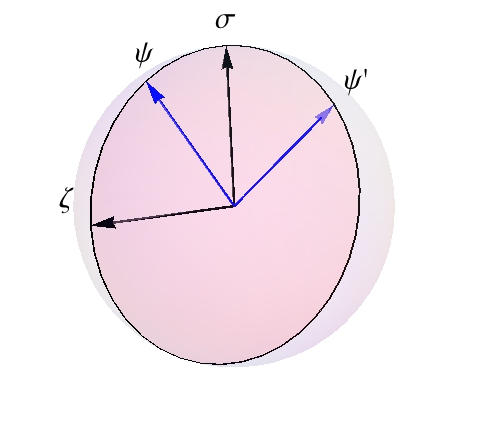}
\caption{\label{fig:psipsip}
Illustration of Lemma 2, with 
$\psi'=|\psi(\frac{ \pi}{2}-\theta,\varphi+\pi) \rangle $
and $\psi=|\psi(\theta,\varphi) \rangle $. }
\end{center}
\end{figure}

\begin{figure}
\begin{center}
\includegraphics[width=6.5cm]{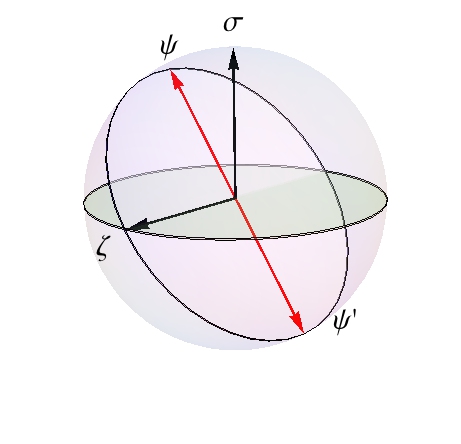}
\caption{\label{fig:psiMinusPsi}
Illustration for Lemma 3, with  $\psi=|\psi(\theta,\varphi) \rangle $ and
$\psi'=|\psi(\pi-\theta,\varphi+\pi) \rangle $.}
\end{center}
\end{figure}

We continue the proof in a similar fashion as Gudder.
Obviously we have $m(\hat{P}_\psi)=1$ if $|\psi \rangle = |\sigma \rangle $. 
If $|\psi \rangle $ is orthogonal to $|\sigma \rangle $ then $\hat{P}_\psi$ is orthogonal to
$\hat{P}_\sigma$. Hence
\begin{align} 
   m(\hat{P}_\sigma+\hat{P}_\psi)&= m(\hat{P}_\sigma) + m(\hat{P}_\psi)
\\ &= 1+  m(\hat{P}_\psi)
\\ &\leq 1 .
\end{align} 
 From this we can infer 
\\[3mm]
{\bf Lemma 1}: if $|\psi \rangle $ is orthogonal
to $| \sigma \rangle $ then  $m(\hat{P}_\psi)=0$.
\\[3mm]
The general state $|\psi(\theta,\varphi) \rangle $ is completely determined
by the two angles $\theta,\varphi$. Because $\hat{P}_\psi$ is in turn
completely specified by the state $|\psi \rangle $, 
we can consider the
function $m$ as a function of these angles, $m(\hat{P}_\psi)
=m(\theta,\varphi)$.  We now derive a set of conditions on this function
of two angles.
\\[3mm]
{\bf Lemma 2}: $m(\frac{ \pi}{2}-\theta,\varphi+\pi) = 1 - m(\theta,\varphi)$.
\\[3mm]
To prove this we refer to Fig.~\ref{fig:psipsip},
where $|\psi(\theta,\varphi) \rangle $ and
$|\psi(\frac{ \pi}{2}-\theta,\varphi+\pi) \rangle $
span a 2D subspace that is also spanned by
$|\sigma \rangle $ and a vector $|\zeta \rangle $ that is orthogonal
to $|\sigma \rangle $. We therefore have
$m(\hat{P}_\zeta)=0$ and $\hat{P}_\psi+\hat{P}_{\psi'} =\hat{P}_\sigma +\hat{P}_\zeta$, so that
\begin{align} 
  m(\hat{P}_\sigma )+m(\hat{P}_\zeta) &= 1 = m(\hat{P}_\psi)+m(\hat{P}_{\psi}'),
\end{align} 
which proves Lemma 2.
\\[3mm]
{\bf Lemma 3}: $m(\pi-\theta, \varphi+\pi)=m(\theta , \varphi)$.
\\[3mm] 
This can be proven by looking at Fig.~\ref{fig:psiMinusPsi},
where $\psi=|\psi(\theta,\varphi) \rangle $ and
$\psi'=|\psi(\pi-\theta,\varphi+\pi) \rangle $. We then have
\begin{align} 
  m(\hat{P}_\psi)+m(\hat{P}_\zeta) &= m(\hat{P}_{\psi'})+m(\hat{P}_\zeta),
\end{align} 
with a vector 
\begin{align} 
  |\zeta \rangle &=  - \sin\varphi |\sigma_\perp \rangle +
     \cos\varphi |\sigma_\perp' \rangle ,
\end{align} 
that is orthogonal to both $| \sigma \rangle $ and 
$|\psi(\theta,\varphi) \rangle $,
so that $m(\hat{P}_\zeta)=0$ $\Box$

Lemma 3 implies that we can restrict our considerations
to angles $0<\theta < \pi/2$. Because of Lemma 2 we can
further reduce this range to $0<\theta < \pi/4$.

\subsection{ $m(\theta,\varphi)$ cannot depend on $\varphi$} 
We now introduce the states
\begin{align} 
  | x \rangle &= \cos \left ( \beta \right )  | \psi(\theta,\varphi)
  \rangle 
   + \sin \left ( \beta \right ) |\zeta \rangle 
\label{eq:xvec}\\
  |y \rangle &=  \sin \left ( \beta \right )
  |\psi(\theta,\varphi)\rangle 
   - \cos \left ( \beta \right )   |\zeta \rangle  \; ,
\end{align} 
which are orthogonal to each other and
span the same 2D subspace as $|\psi(\theta,\varphi) \rangle $ and
$|\zeta \rangle $. We therefore have
\begin{align} 
  \hat{P}_\psi + \hat{P}_\zeta &= \hat{P}_x + \hat{P}_y .
\end{align} 
Because $m(\hat{P}_\zeta)=0$ 
and $\hat{P}_x \hat{P}_y = \hat{P}_\psi \hat{P}_\zeta = 0$, we can conclude that
\begin{align} 
  m(\hat{P}_\psi) &= m(\hat{P}_x) + m(\hat{P}_y).
\label{eq:gudderCentral}\end{align} 
This is a key relation in Gudder's proof, but from this point on we
will deviate from his line of reasoning.

\begin{figure}
\begin{center}
\includegraphics[width=6.5cm]{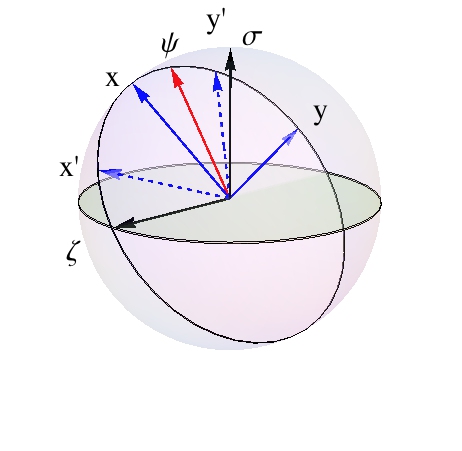}
\caption{\label{fig:gleason}
Sketch of the vectors involved in the derivation of Eq.~(\ref{eq:ThetaThetaX})
for $\beta=\pi/8$.
The horizontal plane corresponds to all vectors orthogonal to 
$|\sigma \rangle $.
The tilted plane corresponds to the plane spanned by
$|\psi(\theta,\varphi) \rangle $ and $|\zeta \rangle $,
or alternatively by $|x \rangle , |y \rangle $ or  $|x' \rangle , |y' \rangle $.}
\end{center}
\end{figure}

The vectors $|x \rangle $, $|y \rangle $ can be expressed in the form
\begin{align} 
  |x\rangle = |\psi(\theta_x,\varphi+\delta\varphi_x)\rangle  
  \; , \;
  |y\rangle = |\psi(\theta_y,\varphi+\delta\varphi_y)\rangle  
\end{align} 
with
\begin{align} 
  \theta_x  &= \arccos\left( \cos\theta  \cos\beta \right)
\label{eq:thetabeta}\\
  \theta_y &= \arccos \left(
   \cos\theta \sin\beta \right) 
\label{eq:thetay}\\
  \delta\varphi_x &= 
  \arctan \left( \csc\theta \tan\beta \right)
\\
  \delta\varphi_y &=
  -\arctan \left(\csc\theta \cot\beta    )\right).
\end{align} 
We can use these vectors for any value of $\beta$, but we are
particularly interested in one arbitrary but fixed value $0<\beta
<\frac{ \pi}{2}$
and a second value $\beta' = \frac{ \pi}{2}-\beta$, which corresponds
to a second orthogonal pair of vectors $|x' \rangle , |y' \rangle $.
A sketch of all of these vectors for $\beta = \pi/8$ is presented in Fig.~\ref{fig:gleason}.
It is not hard to see that 
\begin{align} 
  \theta_{x'}&=\theta_y  \; , \; \theta_{y'}=\theta_x
\\
  \delta\varphi_{x'}&=-\delta\varphi_y  \; , \;
  \delta\varphi_{y'}=-\delta\varphi_x .
\end{align} 
Eq.~(\ref{eq:gudderCentral}) can be evaluated for both pairs $x,y$ and
$x',y'$ of orthogonal vectors so that we arrive at two equations
\begin{align} 
  m(\theta,\varphi) &= m(\theta_x,\varphi +\delta\varphi_x)
  + m(\theta_y,\varphi +\delta\varphi_y) 
\label{eq:gudderCentral2}\\
 m(\theta,\varphi) &= m(\theta_y,\varphi -\delta\varphi_y)
  + m(\theta_x,\varphi -\delta\varphi_x). 
\label{eq:gudderCentral2a}\end{align} 
These equations are valid for all choices of $\varphi$. We can
therefore replace $\varphi$ by $\varphi - \delta \varphi_y$ in
Eq.~(\ref{eq:gudderCentral2}) and by
$\varphi + \delta \varphi_y$ in
Eq.~(\ref{eq:gudderCentral2a}) and then
eliminate $m(\theta_y,\varphi)$ from
Eq.~(\ref{eq:gudderCentral2}) to obtain
\begin{align} 
   m(\theta,\varphi-\delta \varphi_y) &= 
  m(\theta,\varphi+\delta \varphi_y) +
  m(\theta_x,\varphi +\delta\varphi_x-\delta \varphi_y) 
\nonumber \\ & \hspace{4mm} -
 m(\theta_x,\varphi -\delta\varphi_x+\delta \varphi_y) .
\label{eq:ThetaThetaX}\end{align} 

Eq.~(\ref{eq:ThetaThetaX}) is central for our proof because it relates
vectors with overlap $\cos\theta$ (with $|\sigma \rangle $)
to vectors with a different overlap $\cos\theta_x$. 
It will also provide the connection between the proof of Gleason and
that of Kochen and Specker.
For special
values of the angles (e.g., for $\delta \varphi_x=\delta \varphi_y$),
one could use Eq.~(\ref{eq:ThetaThetaX}) to express
$m(\theta_x,\varphi)$ directly in terms of $m(\theta, \ldots)$.
However, to generally achieve such a relation we have to employ 
Fourier transformation.

The function $m(\theta,\varphi)$ is periodic in $\varphi$
and can therefore be expressed as a Fourier series
\begin{align} 
  m(\theta,\varphi) &= \sum_{n=-\infty}^\infty e^{i n \varphi}
  m_n(\theta)
\\
  m_n(\theta) &= \frac{ 1}{2\pi} \int_0^{2\pi}d\varphi\;
  e^{-i n\varphi} m(\theta,\varphi)\;.
\end{align} 
Because $m(\theta,\varphi)$ is real we have the relation $m_{-n}(\theta) = m_n^*(\theta)$.
Taking the Fourier transform of Eq.~(\ref{eq:ThetaThetaX}) and solving
the resulting equation for $m_n(\theta_x)$ yields, for the case $n\neq 0$,
\begin{align} 
  m_n(\theta_x) &= \frac{ \sin \left (n\, \delta\varphi_y\right)}{
   \sin \left (n(\delta\varphi_y-\delta\varphi_x)\right)} m_n(\theta).
\label{eq:mn1}\end{align} 
In the way we derived this equation, the angles $\theta_x,
\delta\varphi_x$ and $\delta\varphi_y$ are functions of an arbitrary
angle $\beta$. However, For $0<\theta \leq \theta_x < \pi/2$, the angle
$\beta$ is uniquely determined by $\theta$ and $\theta_x$ through
$\beta = \text{arccos}(\cos\theta_x / \cos\theta)$, which can be
derived from Eq.~(\ref{eq:thetabeta}). A little algebra with inverse
trigonometric functions then enables us to express 
$\delta\varphi_x$ and $\delta\varphi_y$ through $\theta$ and
$\theta_x$ as
\begin{align} 
   \delta\varphi_y &= - \arctan \left (
  \frac{ \cos\theta_x}{\sin\theta \sqrt{\cos^2\theta-\cos^2\theta_x}}
  \right )
\\
   \delta\varphi_y-\delta\varphi_x  &= \arctan \left (
  \frac{ \sin\theta}{\cos\theta_x \sqrt{\cos^2\theta-\cos^2\theta_x}}
  \right ).
\end{align} 
What we have accomplished in Eq.~(\ref{eq:mn1}) is to establish 
a relation expressing $m_n(\theta_x)$
through $m_n(\theta)$ for an arbitrary pair of angles
$0<\theta \leq \theta_x < \pi/2$. Hence, if we know
$m_n(\theta)$ for one value of $\theta$ we also know
it for angles $\theta_x>\theta$. We are now going to use 
this to show that $m_n(\theta)=0$ for $n\neq 0$.

To do so, we start by considering the Fourier transform of Lemma 2,
which for $n\neq 0$ reads
\begin{align} 
    m_n \left ( \frac{ \pi}{2}-\theta \right ) e^{in\pi} &= -m_n(\theta),
\end{align} 
or $ m_n  ( \frac{ \pi}{2}-\theta ) = (-1)^{n+1}m_n(\theta)$.
On the other hand, if for $\theta \leq \pi/4$ we set $\theta_x =\pi/2
-\theta$ we obtain
\begin{align} 
    \delta\varphi_y-\delta\varphi_x =-\delta\varphi_y =\arctan
    \left (\frac{ 1}{\sqrt{\cos(2\theta)}}\right ),
\end{align} 
so that Eq.~(\ref{eq:mn1}) implies
$ m_n ( \frac{ \pi}{2}-\theta ) = -m_n(\theta)$.
Consequently, Lemma 2 and Eq.~(\ref{eq:mn1}) can both be fulfilled
only if $m_{2n+1}(\theta)=0$.

\begin{figure}
\begin{center}
\includegraphics[width=6.5cm]{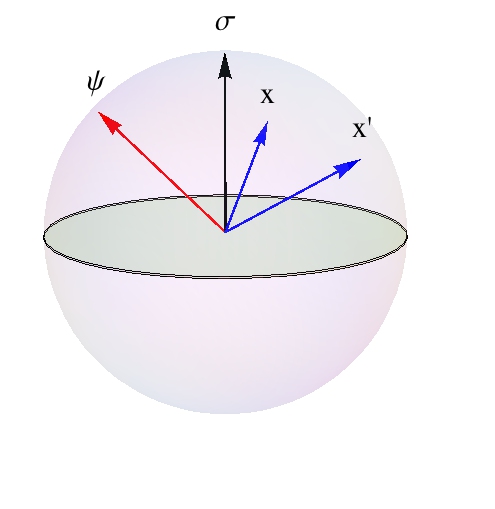}
\caption{\label{fig:psiXXp}
Sketch of the vectors used to show that all even Fourier components
$m_{2n}(\theta)$ must be zero for $n\neq 0$.}
\end{center}
\end{figure}

It remains to show that $m_{2n}(\theta)=0$ as
well. To do so we consider a set of three orthonormal vectors given by
\begin{align} 
  |\psi \rangle &= \left|\psi \left(\frac{ \pi}{4},\varphi\right)\right \rangle 
\\
  |x \rangle &= \left|\psi \left (\frac{ \pi}{3},\varphi+\pi+\arctan \sqrt{2}\right) \right\rangle 
\\
  |x' \rangle &= \left|\psi \left (\frac{ \pi}{3},\varphi+\pi-\arctan \sqrt{2}\right) \right\rangle .
\end{align} 
These vectors are illustrated in Fig.~\ref{fig:psiXXp}.
Because the vectors are orthonormal we have $\hat{P}_\psi+\hat{P}_x+\hat{P}_{x'}=
\mathds{1}$ and therefore
\begin{align} 
  m(\hat{P}_\psi)+m(\hat{P}_x)+m(\hat{P}_{x'}) &= 1.
\end{align} 
For $2n\neq 0$, the even Fourier components of this equation read
\begin{align} 
   m_{2n}\left(\frac{ \pi}{4}\right) +2 \cos(2n \arctan\sqrt{2}) \, 
  m_{2n}\left(\frac{\pi}{3} \right) =0.
\label{eq:finalStep}\end{align} 
For $\theta=\frac{ \pi}{4}$ and $\theta_x=\frac{ \pi}{3}$
we have $\delta\varphi_y = -\arctan \sqrt{2}$ and
 $\delta\varphi_y-\delta\varphi_x = \arctan 2\sqrt{2}$. 
Using Eq.~(\ref{eq:mn1}) to express $m_{2n}(\frac{ \pi}{3})$ through
 $m_{2n}(\frac{ \pi}{4})$ in Eq.~(\ref{eq:finalStep}) we get
\begin{align} 
   2\, m_{2n}\left (\frac{ \pi}{4}\right) &= 0.
\end{align} 
Hence, for $\theta=\pi/4$, all Fourier coefficients $n\neq 0$ are
zero. Because of relation (\ref{eq:mn1}) this also holds for all
angles $\pi/4 \leq \theta \leq \pi/2$. Because of Lemma 2 and 3,
this conclusion must be true for arbitrary values of $\theta$.
We therefore have shown that $m(\theta,\varphi)$ cannot depend on $\varphi$.

\subsection{Determining $m(\theta)$} 
Now that we know that $m$ only depends on $\theta$, relation
(\ref{eq:gudderCentral2}) can be written as 
\begin{align} 
   m(\theta) &= m(\theta_x(\beta))+m(\theta_y(\beta)).
\label{eq:gudderCentral3}\end{align} 
We now make a change of variables from $\theta$ to $u = \cos^2
\theta$, with $\tilde{m}(u) = m(\theta)$. 
Using Eqs.~(\ref{eq:thetabeta}) and (\ref{eq:thetay}),
relation (\ref{eq:gudderCentral3}) can then be written as
\begin{align} 
  \tilde{m}(u) &= \tilde{m}(u \cos^2\beta) 
  +  \tilde{m}(u \sin^2\beta) .
\end{align} 
Setting $u'=u \cos^2\beta$, this can be cast into the form
\begin{align} 
  \tilde{m}(u) &= \tilde{m}(u') 
  +  \tilde{m}(u -u') .
\label{eq:tildemRel}
\end{align} 
We can use this to show that
\begin{align} 
 \tilde{m}(2^{-n}) &= 2^{-n}
\label{eq:contrel1}\\
 \tilde{m}(k u') &= k \tilde{m}( u'),
\label{eq:contrel2}\end{align} 
for $k,n \in \mathds{N}$. To do so, we
set $u'=u/2$ in Eq.~(\ref{eq:tildemRel}) , so that
\begin{align} 
  \tilde{m}(u) &= 2\tilde{m}(u/2). 
\label{eq:powerLaw}\end{align} 
If we now set $u=1$ we get $\tilde{m}(1/2) = 1/2$. Applying relation
(\ref{eq:powerLaw}) $n$ times yields Eq.~(\ref{eq:contrel1}).

To prove Eq.~(\ref{eq:contrel2}) we set $u=k u'$ in
Eq.~(\ref{eq:tildemRel}). We then obtain
\begin{align} 
  \tilde{m}(k u') &= \tilde{m}(u') 
  +  \tilde{m}((k-1)u') .
\label{eq:tildemRel2}
\end{align} 
Because of Eq.~(\ref{eq:powerLaw}),
Eq.~(\ref{eq:contrel2}) is correct for $k=2$. Assuming that it is correct
for $k-1$, Eq.~(\ref{eq:tildemRel2}) yields
\begin{align} 
  \tilde{m}(k u') &= \tilde{m}(u') 
  +  (k-1)\tilde{m}(u') \;  =  \; k \tilde{m}( u').
\end{align} 

By combining Eqs.~(\ref{eq:contrel1}) and (\ref{eq:contrel2}) we
have now shown that $\tilde{m}(u)=u$ for all numbers of the form $u=k 2^{-n}$. 
Now suppose that 
$u'=k_1 2^{-n_1}$ and $u-u'=k_2 2^{-n_2}$. Relation
(\ref{eq:tildemRel}) then implies that this also holds for numbers
of the form $u=k_1 2^{-n_1}+k_2 2^{-n_2}$. By repeating this argument,
we can show that $\tilde{m}(u)=u$ for any number of the form
$u =\sum_r k_r 2^{-r}$. However, this is the binary
representation of real numbers in the interval $[0,1]$, so that
$\tilde{m}(u)=u$ holds for all $u\in [0,1]$. Hence,
\begin{align} 
   m(\theta) = \cos^2\theta
  = \langle \sigma |\hat{P}_\psi|\sigma \rangle .
\label{eq:GproofLast}\end{align} 
This proves the theorem for rank-1 projectors $\hat{P}_\psi$.
Because any projector can be written as a sum of mutually orthogonal
projectors of rank 1, assumption (\ref{eq:assump2}) ensures that the theorem
holds for arbitrary projectors $\Box$ 

\section{The Kochen-Specker Theorem}
The theorem of Kochen and Specker addresses an apparently very
different question. It does not deal with
probabilities but rather asks whether it is possible to assign
specific values to all observables in a system that can be
described using quantum theory. 

Consider the observables corresponding to projectors
on a three-dimensional real Hilbert space. 
In a measurement, all these observables
would take values that are either 0 or 1. In a classical world,
one would expect that observables take their values independently
of whether one actually performs a measurement or not. 
For instance, if we throw a coin and do not look at
  the result, we would still be convinced that it would be either head (0)
  or tail (1).
The 
value of the observables may not be known, but it would appear
plausible that each possible set of values for an observable
could be associated with a certain probability. 
The question is which
sets of values are actually possible, and the answer given by Kochen
and Specker is: none.
In the language used here, their result can be stated
  as follows.
\\[3mm]
{\bf Adapted Kochen-Specker Theorem}: Let ${\cal H}$ be a real 
separable Hilbert space of dimension $\geq 3$ and ${\cal P}({\cal H})$
the lattice of projectors. Then there is no homomorphism 
that maps ${\cal P}({\cal H})$ to the set $\{0,1\}$.
\\[3mm]
In the original theorem, the lattice of projectors is replaced by a
partial Boolean algebra, which includes all observables on ${\cal
  H}$. In our case, the homomorphism is a map 
$m: {\cal P}({\cal H}) \rightarrow \{0,1\}$ that preserves the lattice
structure, i.e., it obeys assumptions (\ref{eq:assump1}) and
(\ref{eq:assump2}) of Gleason's theorem. However, it can only take
the values 0 or 1.

The Kochen-Specker Theorem may be considered 
a corollary of Gleason's theorem: because $m(\hat{P})$ is confined to map a 
projector to the discrete values 0 or 1, Gleason's theorem
tells us that it is impossible because the only possible
map (\ref{eq:GproofLast}) takes continuous values. This connection
between the two theorems is well known and has been used by
Hrushovski and Pitowsky to construct extensions of the 
Kochen-Specker Theorem \cite{Hrushovski2004}.

However, direct proofs of the Kochen-Specker Theorem are much more
intuitive than the proof of Gleason's theorem.
If one associates the values of 0 and 1 with the color blue and red,
respectively, then one has to show
that it is impossible to color the unit sphere (which is formed by the
tips of all unit vectors $|\psi \rangle $) in red and blue in such a
way that (i) all points separated from a red point by a right angle must
be blue, and (ii) that any three points mutually separated by right angles
must contain one red and two blue points. Condition (i) is similar
to the statement of Lemma 1 above: if we know that
$m(\hat{P}_\sigma)=1$ for some vector $|\sigma \rangle $, then $m$
must vanish for all projectors on states that are orthogonal to  $|\sigma \rangle $.
Condition (ii) arises from the fact that, on a three-dimensional
subspace of ${\cal H}$, we have $m(\mathds{1})=1=m(\hat{P}_1)+
m(\hat{P}_2)+m(\hat{P}_3)$
for three orthogonal rank-1 projectors $\hat{P}_i$.

Kochen and Specker constructed a set of 117 vectors for which
no consistent choice of colours could be made. The theorem has later
been derived for larger Hilbert spaces and with fewer basis vectors
\cite{PhysRevLett.65.3373,0305-4470-24-4-003,Kernaghan19951,Cabello1996183,PhysRevLett.108.030402,PhysRevA.88.012102,Pavicic20102122,Toh20104834},
and has been generalized to open quantum systems
\cite{PhysRevLett.90.190401,PhysRevA.68.052104}.

Despite being a corollary of Gleason's theorem,
the Kochen-Specker theorem makes a stronger statement
about contextual HV theories. The reason is that
in Gleason's theorem $m$ represents a probability distribution, while
in the Kochen-Specker theorem $m$ represents 
the allowed measurement values. One can distinguish
two types of contextual HV theories: type I only allows the
probability distribution to be context-dependent, while type II
admits the possibility that both probability distribution and
measurement values may depend on the experimental context.

The contextual measurement values in type II introduce a new challenge.
Suppose we want to measure the sum $\hat{P}_x + \hat{P}_z$
of two non-commuting projectors. In each run of the experiment
we would have to add the values measured for both observables,
but since they cannot be measured simultaneously, this is 
not possible in practice. 
However, if quantum theory could be interpreted in
terms of HV theories, both observables would
need to take some value, regardless of whether we can actually
measure it. One therefore had to introduce 
counterfactual values
\cite{PhysRevD.3.1303,Eberhard1977,Redhead:Incompleteness}
that an observable would take even if the experiment is not set
up to measure it.

Counterfactual values can be avoided if contextuality is defined in
a refined way. Suppose $\hat{A}$ is an observable that commutes with two
other observables $\hat{B}$ and $\hat{C}$, but 
$[\hat{B},\hat{C}] \neq 0$.
In this case we can simultaneously measure $\hat{A}$ and $\hat{B}$, or
 $\hat{A}$ and $\hat{C}$. Then observable $\hat{A}$ is non-contextual
if the measurement outcomes do not depend on whether it is measured
simultaneously with $\hat{B}$ or $\hat{C}$ \cite{PhysRevA.89.032109}. This definition is
well suited for projection measurements in quantum theory, but has been
generalized by Spekkens \cite{PhysRevA.71.052108} to include 
unsharp measurements and more general physical models.

Gleason's theorem can neither exclude type I nor type II. 
The Kochen-Specker theorem can exclude type I because it
does not depend on the probability 
distribution. 
Bell inequalities
\cite{Bell:Physics1964,RevModPhys.38.447,PhysRevLett.23.880} 
can exclude 
``local'' HV models of type I, where the probability distribution
can only depend on local changes of the apparatus, not on distant changes 
that would require superluminal speed to affect the
probability distribution \cite{PhysRevLett.48.291}. 

We proceed by using the methods of Sec.~\ref{sec:GleasonProof}
prove the Kochen-Specker theorem on a real three-dimensional subspace
of ${\cal H}$. If there can be no homomorphism for this subspace, then
there can also be no homomorphism on ${\cal H}$. Because of condition
(ii) there must be at least one red point on the unit sphere, which we
call $|\sigma \rangle $. Without loss of generality, we put this point
at the North pole of the sphere.
Lemma 1 then
ensures that all points on the equator must be blue. Lemma 2
implies that if $|\psi(\theta ,\varphi) \rangle $ is red then
$|\psi' \rangle = |\psi(\frac{ \pi}{2}-\theta ,\varphi+\pi) \rangle$
must be blue, or vice versa.

Turning to Fig.~\ref{fig:gleason} we can see that relation
(\ref{eq:gudderCentral2}) connects the colours of $|\psi \rangle $,
$|x \rangle $ and $|y \rangle $. If $|\psi \rangle $ is red, then
one of $|x \rangle $ and $|y \rangle $ must be red.
If $|\psi \rangle $ is blue, then
both $|x \rangle $ and $|y \rangle $ must be blue as well. Because
the choice of $|x \rangle $ and $|y \rangle $ is arbitrary, the entire
plane spanned by $|\psi \rangle $ and $|\zeta \rangle $ must then be blue.

\begin{figure}
\begin{center}
a)\includegraphics[width=6.5cm]{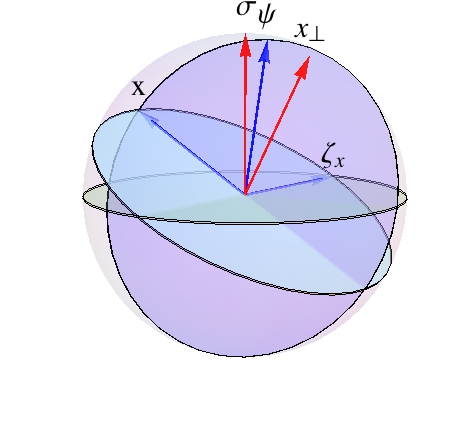}
\\
b)\includegraphics[width=6.5cm]{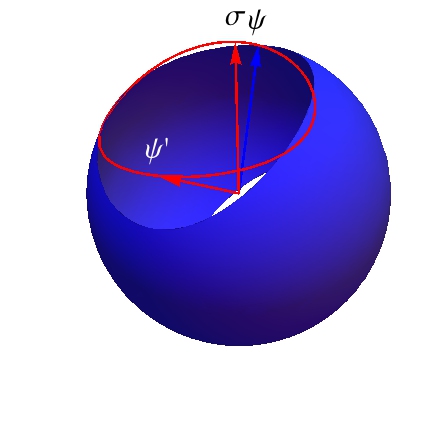}
\caption{\label{fig:KS1}
Sketch for a proof of the Kochen-Specker theorem for the choice
$\theta=\pi/10$. a) 
If $\psi$ is blue then the three disks must be blue as well.
b) The blue area is the coverage of all disks spanned by $x$ and $\zeta_x$ for all
choices of $\beta$. The red line corresponds to vectors that must be red. }
\end{center}
\end{figure}

This observation enables us to construct a contradiction: suppose
 $|\psi \rangle $ is blue. We then know that both the equator and the
plane spanned by $|\psi \rangle $ and $|\zeta \rangle $ must be blue.
For a given vector $|x \rangle $ of Eq.~(\ref{eq:xvec}), we can find
another vector
\begin{align} 
  |\zeta_x \rangle &=\frac{ \sin\theta}{\sqrt{1+\cot^2\beta
      \sin^2\theta}}
  \left (|\psi \rangle -\cot\beta \, |\zeta \rangle \right )
\end{align} 
that lies on the equator and is orthogonal to $|x \rangle $.
The two vectors  $|x \rangle $ and $|\zeta_x \rangle$ are both blue
and therefore span a plane that must be blue. This is depicted in
Fig.~\ref{fig:KS1}~a): if $|\sigma \rangle $ is red and $|\psi \rangle
$ is blue, then the equator, the plane spanned by $|\psi \rangle $ and $|x \rangle
$, and the plane spanned by $|x \rangle $ and $|\zeta_x \rangle$ must
all be blue. Furthermore the vector $|x_\perp \rangle = |\zeta_x
\rangle \times |x \rangle $ must be red, where $\times$ denotes the
vector cross product in three dimensions.

Fig.~\ref{fig:KS1} a) shows the blue plane spanned by
$|x \rangle $ and $|\zeta_x \rangle$  for one particular choice
of $|x \rangle $. However, by varying $\beta $ in Eq.~(\ref{eq:xvec}) we can
continuously change this plane from the plane
spanned by $|\psi \rangle $ and $|\zeta \rangle $ (for $\beta=0$)
into the equatorial plane (for $\beta=\pi/2$ ). The set of all points
that lie on any of these planes forms a blue area on the unit sphere
that is shown in Fig.~\ref{fig:KS1} b). Each of the planes also
determines a red vector $|x_\perp \rangle $. As $\beta$ varies, 
this vector moves along a trajectory connecting 
$|\psi' \rangle $  (for $\beta=0$) with $|\sigma \rangle $
(for $\beta=\pi/2$ ). This trajectory is shown in red in
Fig.~\ref{fig:KS1} b) for $\theta=\pi/10$.

The size of the blue area and the red trajectory depends on the
angle $\theta$ between $|\psi \rangle $ and $|\sigma \rangle $. 
For values of $\theta \approx \pi/2$, the blue area essentially 
corresponds to a blue ribbon around the equator and the red 
trajectory stays close to the north pole. However, the size of both
the area and the trajectory grows as $\theta$ shrinks and they start
to overlap for values of about $\theta \leq \pi/10$. 
More precisely, we have numerically determined that the maximal angle
for which the red trajectory and the blue area are overlapping is
$\theta \approx 0.108 \,\pi$; we conjecture that the precise boundary
is at $\theta =\text{arccos}(\sqrt{8/9})$.

As no point
can be both red and blue, the assumption that $|\psi \rangle $
is blue must therefore be wrong for $\theta \leq \pi/10$.
Because we have not made any assumption about $|\sigma \rangle $
apart from being red, we have just shown that any point that is closer
than $\pi/10$ from a red point must also be red. By repeated
application of this principle to different red points
we can infer that the entire unit sphere must be red, which would be
in contradiction with the assumptions $\Box$

The above argument uses an infinite number of vectors (the blue area
of the sphere), but it can easily be reduced to a finite number of
vectors. For simplicity we consider a vector $|\psi (\theta,\varphi)\rangle $ of
Eq.~(\ref{eq:psiGeneral}) for which $\theta=\pi/10$ and $\varphi=0$.
We then can pick one pair of vectors $|x \rangle , |\zeta_x \rangle$,
characterized by an angle $\beta$, for which $|x_\perp \rangle $ is red at a
specific point, and a second one, characterized by an angle $\beta'$,
for which this point lies on the plane spanned by the pair and hence
must be blue. To be specific, we
numerically determined $\beta\approx 0.756 \pi$,
for which $|x_\perp \rangle \approx |\psi(0.24\pi, 0.599\pi)\rangle $, and 
 $\beta'\approx 0.137 \pi$.
Because this leads to a contradiction, we
 can infer as before that $|\psi (\frac{ \pi}{10},0) \rangle $ must be red. We then can
 repeat this procedure to show that all vectors 
 $|\psi (n \frac{ \pi}{10},0) \rangle $, with $n=0,1,2,\cdots$ must be
 red. However, for $n=5$ this vector is located on the equator and
 thus has to be blue, which proves the adapted Kochen-Specker theorem
by using a finite set of vectors only. This argument
is close to the original proof of the theorem.

\section{Discussion and Conclusion}
In this paper we have provided alternative proofs to a reduced version of Gleason's theorem
and the Kochen-Specker theorem. Both theorems are concerned
with a function $m(\hat{P})$ that maps the lattice of projectors on a
real Hilbert space ${\cal H}$ to real numbers.
The main difference is the image of $m$, which is given by $[0,1]$ for
Gleason's theorem and $\{0,1\}$ for the Kochen-Specker theorem.

In our approach, both proofs utilize Lemma 1, Lemma 2, and 
Eq.~(\ref{eq:gudderCentral2}), which establish algebraic
relations between the values of $m$ for orthogonal 
projectors or, equivalently, orthogonal unit vectors. 
In particular, Eq.~(\ref{eq:gudderCentral2}) relates 
$m$ for sets of unit vectors that are rotated around the vertical
axis.

Mathematically, Gleason's theorem is the stronger result and 
normally requires more powerful techniques for its proof. The 
methods developed here give further insight into this.
It is well known that the Kochen-Specker theorem can be proven using
only a finite number of unit vectors. To prove the reduced
Gleason theorem, all vectors on the unit sphere are required.
Technically, one may say that the Kochen-Specker theorem
employs Eq.~(\ref{eq:gudderCentral2}) for a finite number of
discrete values of the rotation angle, while Gleason's theorem requires 
continuous values. One can then use Fourier transformation to solve
the corresponding algebraic relations.

Apart from providing a new perspective on the relation between both
theorems, the techniques developed here may also be useful for
extensions of the Kochen-Specker theorem that do not require the
use of Gleason's theorem. For instance, one may be able to find a set
of observables that take more discrete values than just 0 and 1
and can be related by finite set of rotation angles in
Eq.~(\ref{eq:gudderCentral2}), but this is beyond the scope of this paper.

\acknowledgements 
We thank James Malley for helpful comments on a previous version
of the manuscript and the Natural Sciences and Engineering Research Council of Canada (NSERC) for financial support.

\begin{appendix}
\section{Projector lattices}\label{app:projectors}
On a finite-dimensional Hilbert space ${\cal H}$, 
a projector $\hat{P}$ corresponds to a matrix
that satisfies $\hat{P}^2 =\hat{P}$ and only has eigenvalues 0 and 1.
It maps a state $|\psi \rangle  \in {\cal H}$ to that part
$\hat{P}|\psi \rangle $ of the state that lies in
a given subspace of  ${\cal H}$.
Rank-1 projectors take the form $\hat{P}_\phi = |\phi \rangle \langle
\phi|$ for some unit vector $|\phi \rangle $ and
project on one-dimensional subspaces.

Projectors can be added, but the sum of two (or more) projectors is
only a projector if they project on mutually
orthogonal subspaces. The set of all projectors on a given Hilbert
space, together with the rules how to add them to get new projectors,
is called the lattice of projectors. More details can be found in 
Ref.~\cite{Beltrametti:QM81}, for instance.

\end{appendix}

\bibliographystyle{apsrev4-1}
\bibliography{/Users/pmarzlin/Documents/literatur/kpmJabRef.bib}

\begin{thebibliography}{44}%
\makeatletter
\providecommand \@ifxundefined [1]{%
 \@ifx{#1\undefined}
}%
\providecommand \@ifnum [1]{%
 \ifnum #1\expandafter \@firstoftwo
 \else \expandafter \@secondoftwo
 \fi
}%
\providecommand \@ifx [1]{%
 \ifx #1\expandafter \@firstoftwo
 \else \expandafter \@secondoftwo
 \fi
}%
\providecommand \natexlab [1]{#1}%
\providecommand \enquote  [1]{``#1''}%
\providecommand \bibnamefont  [1]{#1}%
\providecommand \bibfnamefont [1]{#1}%
\providecommand \citenamefont [1]{#1}%
\providecommand \href@noop [0]{\@secondoftwo}%
\providecommand \href [0]{\begingroup \@sanitize@url \@href}%
\providecommand \@href[1]{\@@startlink{#1}\@@href}%
\providecommand \@@href[1]{\endgroup#1\@@endlink}%
\providecommand \@sanitize@url [0]{\catcode `\\12\catcode `\$12\catcode
  `\&12\catcode `\#12\catcode `\^12\catcode `\_12\catcode `\%12\relax}%
\providecommand \@@startlink[1]{}%
\providecommand \@@endlink[0]{}%
\providecommand \url  [0]{\begingroup\@sanitize@url \@url }%
\providecommand \@url [1]{\endgroup\@href {#1}{\urlprefix }}%
\providecommand \urlprefix  [0]{URL }%
\providecommand \Eprint [0]{\href }%
\providecommand \doibase [0]{http://dx.doi.org/}%
\providecommand \selectlanguage [0]{\@gobble}%
\providecommand \bibinfo  [0]{\@secondoftwo}%
\providecommand \bibfield  [0]{\@secondoftwo}%
\providecommand \translation [1]{[#1]}%
\providecommand \BibitemOpen [0]{}%
\providecommand \bibitemStop [0]{}%
\providecommand \bibitemNoStop [0]{.\EOS\space}%
\providecommand \EOS [0]{\spacefactor3000\relax}%
\providecommand \BibitemShut  [1]{\csname bibitem#1\endcsname}%
\let\auto@bib@innerbib\@empty
\bibitem [{\citenamefont {Gleason}(1957)}]{GleasonsTheorem}%
  \BibitemOpen
  \bibfield  {author} {\bibinfo {author} {\bibfnamefont {A.~M.}\ \bibnamefont
  {Gleason}},\ }\href {http://www.iumj.indiana.edu/oai/1957/6/56050/56050.xml}
  {\bibfield  {journal} {\bibinfo  {journal} {Indiana Univ. Math. J.}\ }\textbf
  {\bibinfo {volume} {6}},\ \bibinfo {pages} {885} (\bibinfo {year}
  {1957})}\BibitemShut {NoStop}%
\bibitem [{\citenamefont {Kochen}\ and\ \citenamefont
  {Specker}(1968)}]{KochenSpecker68}%
  \BibitemOpen
  \bibfield  {author} {\bibinfo {author} {\bibfnamefont {S.}~\bibnamefont
  {Kochen}}\ and\ \bibinfo {author} {\bibfnamefont {E.}~\bibnamefont
  {Specker}},\ }\href {http://www.iumj.indiana.edu/oai/1968/17/17004/17004.xml}
  {\bibfield  {journal} {\bibinfo  {journal} {Indiana Univ. Math. J.}\ }\textbf
  {\bibinfo {volume} {17}},\ \bibinfo {pages} {59} (\bibinfo {year}
  {1968})}\BibitemShut {NoStop}%
\bibitem [{\citenamefont {Redhead}(1987)}]{Redhead:Incompleteness}%
  \BibitemOpen
  \bibfield  {author} {\bibinfo {author} {\bibfnamefont {M.}~\bibnamefont
  {Redhead}},\ }\href@noop {} {\emph {\bibinfo {title} {Incompleteness,
  Nonlocality and Realism}}}\ (\bibinfo  {publisher} {Clarendon},\ \bibinfo
  {year} {1987})\BibitemShut {NoStop}%
\bibitem [{\citenamefont {Hemmick}\ and\ \citenamefont
  {Shakur}(2012)}]{Hemmick2012}%
  \BibitemOpen
  \bibfield  {author} {\bibinfo {author} {\bibfnamefont {D.~L.}\ \bibnamefont
  {Hemmick}}\ and\ \bibinfo {author} {\bibfnamefont {A.~M.}\ \bibnamefont
  {Shakur}},\ }\href@noop {} {\emph {\bibinfo {title} {Bell's Theorem and
  Quantum Realism}}}\ (\bibinfo  {publisher} {Springer},\ \bibinfo {address}
  {Heidelberg},\ \bibinfo {year} {2012})\BibitemShut {NoStop}%
\bibitem [{\citenamefont {Granstr\"om}(2006)}]{Granstrom2006}%
  \BibitemOpen
  \bibfield  {author} {\bibinfo {author} {\bibfnamefont {H.}~\bibnamefont
  {Granstr\"om}},\ }\emph {\bibinfo {title} {Gleason's theorem}},\ \href
  {http://kof.physto.se/cond_mat_page/theses/helena-master.pdf} {Master's
  thesis},\ \bibinfo  {school} {Stockholm University} (\bibinfo {year}
  {2006})\BibitemShut {NoStop}%
\bibitem [{\citenamefont {Bengtsson}(2012)}]{Bengtsson:GleasonKS}%
  \BibitemOpen
  \bibfield  {author} {\bibinfo {author} {\bibfnamefont {I.}~\bibnamefont
  {Bengtsson}},\ }\href {http://arxiv.org/abs/1210.0436} {\bibfield  {journal}
  {\bibinfo  {journal} {arXiv}\ }\textbf {\bibinfo {volume} {1210.0436}},\
  \bibinfo {pages} {8} (\bibinfo {year} {2012})}\BibitemShut {NoStop}%
\bibitem [{\citenamefont {Cooke}\ \emph {et~al.}(1985)\citenamefont {Cooke},
  \citenamefont {Keane},\ and\ \citenamefont {Moran}}]{PSP:2091976}%
  \BibitemOpen
  \bibfield  {author} {\bibinfo {author} {\bibfnamefont {R.}~\bibnamefont
  {Cooke}}, \bibinfo {author} {\bibfnamefont {M.}~\bibnamefont {Keane}}, \ and\
  \bibinfo {author} {\bibfnamefont {W.}~\bibnamefont {Moran}},\ }\href
  {http://journals.cambridge.org/article_S0305004100063313} {\bibfield
  {journal} {\bibinfo  {journal} {Math. Proc. Camb. Philos. Soc.}\ }\textbf
  {\bibinfo {volume} {98}},\ \bibinfo {pages} {117} (\bibinfo {year}
  {1985})}\BibitemShut {NoStop}%
\bibitem [{\citenamefont {Hellman}(1993)}]{hellmanGleason1993}%
  \BibitemOpen
  \bibfield  {author} {\bibinfo {author} {\bibfnamefont {G.}~\bibnamefont
  {Hellman}},\ }\href {http://dx.doi.org/10.1007/BF01049261} {\bibfield
  {journal} {\bibinfo  {journal} {J. Philos. Logic}\ }\textbf {\bibinfo
  {volume} {22}},\ \bibinfo {pages} {193} (\bibinfo {year} {1993})}\BibitemShut
  {NoStop}%
\bibitem [{\citenamefont {Billinge}(1997)}]{billingeGleason1997}%
  \BibitemOpen
  \bibfield  {author} {\bibinfo {author} {\bibfnamefont {H.}~\bibnamefont
  {Billinge}},\ }\href {http://dx.doi.org/10.1023/A%3A1004275113665} {\bibfield
   {journal} {\bibinfo  {journal} {J. Philos. Logic}\ }\textbf {\bibinfo
  {volume} {26}},\ \bibinfo {pages} {661} (\bibinfo {year} {1997})}\BibitemShut
  {NoStop}%
\bibitem [{\citenamefont {Richman}\ and\ \citenamefont
  {Bridges}(1999)}]{Richman1999287}%
  \BibitemOpen
  \bibfield  {author} {\bibinfo {author} {\bibfnamefont {F.}~\bibnamefont
  {Richman}}\ and\ \bibinfo {author} {\bibfnamefont {D.}~\bibnamefont
  {Bridges}},\ }\href
  {http://www.sciencedirect.com/science/article/pii/S0022123698933729}
  {\bibfield  {journal} {\bibinfo  {journal} {J. Func. Analysis}\ }\textbf
  {\bibinfo {volume} {162}},\ \bibinfo {pages} {287 } (\bibinfo {year}
  {1999})}\BibitemShut {NoStop}%
\bibitem [{\citenamefont {Richman}(2000)}]{RichmanGleason2000}%
  \BibitemOpen
  \bibfield  {author} {\bibinfo {author} {\bibfnamefont {F.}~\bibnamefont
  {Richman}},\ }\href {http://dx.doi.org/10.1023/A%3A1004791723301} {\bibfield
  {journal} {\bibinfo  {journal} {J. Philos. Logic}\ }\textbf {\bibinfo
  {volume} {29}},\ \bibinfo {pages} {425} (\bibinfo {year} {2000})}\BibitemShut
  {NoStop}%
\bibitem [{\citenamefont {Isham}\ \emph {et~al.}(1994)\citenamefont {Isham},
  \citenamefont {Linden},\ and\ \citenamefont
  {Schreckenberg}}]{jmp/35/12/10.1063/1.530679}%
  \BibitemOpen
  \bibfield  {author} {\bibinfo {author} {\bibfnamefont {C.~J.}\ \bibnamefont
  {Isham}}, \bibinfo {author} {\bibfnamefont {N.}~\bibnamefont {Linden}}, \
  and\ \bibinfo {author} {\bibfnamefont {S.}~\bibnamefont {Schreckenberg}},\
  }\href
  {http://scitation.aip.org/content/aip/journal/jmp/35/12/10.1063/1.530679}
  {\bibfield  {journal} {\bibinfo  {journal} {J. Math. Phys.}\ }\textbf
  {\bibinfo {volume} {35}},\ \bibinfo {pages} {6360} (\bibinfo {year}
  {1994})}\BibitemShut {NoStop}%
\bibitem [{\citenamefont {Busch}(2003)}]{PhysRevLett.91.120403}%
  \BibitemOpen
  \bibfield  {author} {\bibinfo {author} {\bibfnamefont {P.}~\bibnamefont
  {Busch}},\ }\href {http://link.aps.org/doi/10.1103/PhysRevLett.91.120403}
  {\bibfield  {journal} {\bibinfo  {journal} {Phys. Rev. Lett.}\ }\textbf
  {\bibinfo {volume} {91}},\ \bibinfo {pages} {120403} (\bibinfo {year}
  {2003})}\BibitemShut {NoStop}%
\bibitem [{\citenamefont {Caves}\ \emph {et~al.}(2004)\citenamefont {Caves},
  \citenamefont {Fuchs}, \citenamefont {Manne},\ and\ \citenamefont
  {Renes}}]{CavesGleason2004}%
  \BibitemOpen
  \bibfield  {author} {\bibinfo {author} {\bibfnamefont {C.}~\bibnamefont
  {Caves}}, \bibinfo {author} {\bibfnamefont {C.}~\bibnamefont {Fuchs}},
  \bibinfo {author} {\bibfnamefont {K.}~\bibnamefont {Manne}}, \ and\ \bibinfo
  {author} {\bibfnamefont {J.}~\bibnamefont {Renes}},\ }\href
  {http://dx.doi.org/10.1023/B%3AFOOP.0000019581.00318.a5} {\bibfield
  {journal} {\bibinfo  {journal} {Found. Phys.}\ }\textbf {\bibinfo {volume}
  {34}},\ \bibinfo {pages} {193} (\bibinfo {year} {2004})}\BibitemShut
  {NoStop}%
\bibitem [{\citenamefont {Edwards}\ and\ \citenamefont
  {R\"uttimann}(1999)}]{WstarGleason}%
  \BibitemOpen
  \bibfield  {author} {\bibinfo {author} {\bibfnamefont {C.~M.}\ \bibnamefont
  {Edwards}}\ and\ \bibinfo {author} {\bibfnamefont {G.~T.}\ \bibnamefont
  {R\"uttimann}},\ }\href {http://dx.doi.org/10.1007/s002200050612} {\bibfield
  {journal} {\bibinfo  {journal} {Comm. Math. Phys.}\ }\textbf {\bibinfo
  {volume} {203}},\ \bibinfo {pages} {269} (\bibinfo {year}
  {1999})}\BibitemShut {NoStop}%
\bibitem [{\citenamefont {Edalat}(2004)}]{Edalat2004}%
  \BibitemOpen
  \bibfield  {author} {\bibinfo {author} {\bibfnamefont {A.}~\bibnamefont
  {Edalat}},\ }\href {http://dx.doi.org/10.1023/B%3AIJTP.0000048823.93080.7e}
  {\bibfield  {journal} {\bibinfo  {journal} {Int. J. Theor. Phys.}\ }\textbf
  {\bibinfo {volume} {43}},\ \bibinfo {pages} {1827} (\bibinfo {year}
  {2004})}\BibitemShut {NoStop}%
\bibitem [{\citenamefont {Wallach}(2000)}]{Wallach0050028}%
  \BibitemOpen
  \bibfield  {author} {\bibinfo {author} {\bibfnamefont {N.~R.}\ \bibnamefont
  {Wallach}},\ }\href {http://arxiv.org/abs/quant-ph/0002058v1} {\bibfield
  {journal} {\bibinfo  {journal} {arXiv.org}\ }\textbf {\bibinfo {volume}
  {quant-ph/0050028}},\ \bibinfo {pages} {9} (\bibinfo {year}
  {2000})}\BibitemShut {NoStop}%
\bibitem [{\citenamefont {Gudder}(1979)}]{GudderQM1979}%
  \BibitemOpen
  \bibfield  {author} {\bibinfo {author} {\bibfnamefont {S.}~\bibnamefont
  {Gudder}},\ }\href@noop {} {\emph {\bibinfo {title} {Stochastic Methods in
  Quantum Mechanics}}}\ (\bibinfo  {publisher} {North-Holland},\ \bibinfo
  {address} {New York},\ \bibinfo {year} {1979})\BibitemShut {NoStop}%
\bibitem [{\citenamefont {Bell}(1964)}]{Bell:Physics1964}%
  \BibitemOpen
  \bibfield  {author} {\bibinfo {author} {\bibfnamefont {J.~S.}\ \bibnamefont
  {Bell}},\ }\href
  {http://philoscience.unibe.ch/documents/TexteHS10/bell1964epr.pdf} {\bibfield
   {journal} {\bibinfo  {journal} {Physics}\ }\textbf {\bibinfo {volume} {1}},\
  \bibinfo {pages} {195} (\bibinfo {year} {1964})}\BibitemShut {NoStop}%
\bibitem [{\citenamefont {Bell}(1966)}]{RevModPhys.38.447}%
  \BibitemOpen
  \bibfield  {author} {\bibinfo {author} {\bibfnamefont {J.~S.}\ \bibnamefont
  {Bell}},\ }\href
  {http://journals.aps.org/rmp/abstract/10.1103/RevModPhys.38.447} {\bibfield
  {journal} {\bibinfo  {journal} {Rev. Mod. Phys.}\ }\textbf {\bibinfo {volume}
  {38}},\ \bibinfo {pages} {447} (\bibinfo {year} {1966})}\BibitemShut
  {NoStop}%
\bibitem [{\citenamefont {Saglamyurek}\ \emph {et~al.}(2011)\citenamefont
  {Saglamyurek}, \citenamefont {Sinclair}, \citenamefont {Jin}, \citenamefont
  {Slater}, \citenamefont {Oblak}, \citenamefont {Bussières}, \citenamefont
  {George}, \citenamefont {Ricken}, \citenamefont {Sohler},\ and\ \citenamefont
  {Tittel}}]{Tittel2011:Nature09719}%
  \BibitemOpen
  \bibfield  {author} {\bibinfo {author} {\bibfnamefont {E.}~\bibnamefont
  {Saglamyurek}}, \bibinfo {author} {\bibfnamefont {N.}~\bibnamefont
  {Sinclair}}, \bibinfo {author} {\bibfnamefont {J.}~\bibnamefont {Jin}},
  \bibinfo {author} {\bibfnamefont {J.~A.}\ \bibnamefont {Slater}}, \bibinfo
  {author} {\bibfnamefont {D.}~\bibnamefont {Oblak}}, \bibinfo {author}
  {\bibfnamefont {F.}~\bibnamefont {Bussières}}, \bibinfo {author}
  {\bibfnamefont {M.}~\bibnamefont {George}}, \bibinfo {author} {\bibfnamefont
  {R.}~\bibnamefont {Ricken}}, \bibinfo {author} {\bibfnamefont
  {W.}~\bibnamefont {Sohler}}, \ and\ \bibinfo {author} {\bibfnamefont
  {W.}~\bibnamefont {Tittel}},\ }\href {\doibase 10.1038/nature09719}
  {\bibfield  {journal} {\bibinfo  {journal} {Nature}\ }\textbf {\bibinfo
  {volume} {469}},\ \bibinfo {pages} {512–515} (\bibinfo {year}
  {2011})}\BibitemShut {NoStop}%
\bibitem [{\citenamefont {Szabo}\ and\ \citenamefont
  {Ostlund}(1996)}]{SzaboOstlund1996}%
  \BibitemOpen
  \bibfield  {author} {\bibinfo {author} {\bibfnamefont {A.}~\bibnamefont
  {Szabo}}\ and\ \bibinfo {author} {\bibfnamefont {N.~S.}\ \bibnamefont
  {Ostlund}},\ }\href@noop {} {\emph {\bibinfo {title} {Modern Quantum
  Chemistry}}}\ (\bibinfo  {publisher} {Dover Publications},\ \bibinfo {year}
  {1996})\BibitemShut {NoStop}%
\bibitem [{\citenamefont {Malley}(2004)}]{PhysRevA.69.022118}%
  \BibitemOpen
  \bibfield  {author} {\bibinfo {author} {\bibfnamefont {J.~D.}\ \bibnamefont
  {Malley}},\ }\href {\doibase 10.1103/PhysRevA.69.022118} {\bibfield
  {journal} {\bibinfo  {journal} {Phys. Rev. A}\ }\textbf {\bibinfo {volume}
  {69}},\ \bibinfo {pages} {022118} (\bibinfo {year} {2004})}\BibitemShut
  {NoStop}%
\bibitem [{\citenamefont {Marzlin}\ and\ \citenamefont
  {Osborn}(2014)}]{PhysRevA.89.032123}%
  \BibitemOpen
  \bibfield  {author} {\bibinfo {author} {\bibfnamefont {K.-P.}\ \bibnamefont
  {Marzlin}}\ and\ \bibinfo {author} {\bibfnamefont {T.~A.}\ \bibnamefont
  {Osborn}},\ }\href {\doibase 10.1103/PhysRevA.89.032123} {\bibfield
  {journal} {\bibinfo  {journal} {Phys. Rev. A}\ }\textbf {\bibinfo {volume}
  {89}},\ \bibinfo {pages} {032123} (\bibinfo {year} {2014})}\BibitemShut
  {NoStop}%
\bibitem [{\citenamefont {Moretti}\ and\ \citenamefont
  {Pastorello}(2013)}]{MorettiGleasonSphericalHarmonics}%
  \BibitemOpen
  \bibfield  {author} {\bibinfo {author} {\bibfnamefont {V.}~\bibnamefont
  {Moretti}}\ and\ \bibinfo {author} {\bibfnamefont {D.}~\bibnamefont
  {Pastorello}},\ }\href {http://dx.doi.org/10.1007/s00023-012-0220-x}
  {\bibfield  {journal} {\bibinfo  {journal} {Ann. H. Poincar\'e}\ }\textbf
  {\bibinfo {volume} {14}},\ \bibinfo {pages} {1435} (\bibinfo {year}
  {2013})}\BibitemShut {NoStop}%
\bibitem [{Note1()}]{Note1}%
  \BibitemOpen
  \bibinfo {note} {There is a counterexample for Gleason's theorem and the
  Kochen-Specker theorem in the two-dimensional case \cite
  {Redhead:Incompleteness}, where $m(\protect \mathaccentV {hat}05E{P})$ can be
  considered as a function $m(\varphi )$ of the angle $\varphi $ on the unit
  circle. The choice $m(\varphi ) = 0$ for $\varphi \in [0, \protect \frac {
  \pi }{2}) \cup [\pi , \protect \frac { 3\pi }{2})$ and $m(\varphi ) = 1$
  elsewhere then fulfills assumptions (\ref {eq:assump1}) and (\ref
  {eq:assump2}) but contradicts the statement of both theorems.}\BibitemShut
  {Stop}%
\bibitem [{\citenamefont {Hrushovski}\ and\ \citenamefont
  {Pitowsky}(2004)}]{Hrushovski2004}%
  \BibitemOpen
  \bibfield  {author} {\bibinfo {author} {\bibfnamefont {E.}~\bibnamefont
  {Hrushovski}}\ and\ \bibinfo {author} {\bibfnamefont {I.}~\bibnamefont
  {Pitowsky}},\ }\href {\doibase 10.1016/j.shpsb.2003.10.002} {\bibfield
  {journal} {\bibinfo  {journal} {Stud. Hist. Phil. Sci. B}\ }\textbf {\bibinfo
  {volume} {35}},\ \bibinfo {pages} {177–194} (\bibinfo {year}
  {2004})}\BibitemShut {NoStop}%
\bibitem [{\citenamefont {Mermin}(1990)}]{PhysRevLett.65.3373}%
  \BibitemOpen
  \bibfield  {author} {\bibinfo {author} {\bibfnamefont {N.~D.}\ \bibnamefont
  {Mermin}},\ }\href {http://link.aps.org/doi/10.1103/PhysRevLett.65.3373}
  {\bibfield  {journal} {\bibinfo  {journal} {Phys. Rev. Lett.}\ }\textbf
  {\bibinfo {volume} {65}},\ \bibinfo {pages} {3373} (\bibinfo {year}
  {1990})}\BibitemShut {NoStop}%
\bibitem [{\citenamefont {Peres}(1991)}]{0305-4470-24-4-003}%
  \BibitemOpen
  \bibfield  {author} {\bibinfo {author} {\bibfnamefont {A.}~\bibnamefont
  {Peres}},\ }\href {http://stacks.iop.org/0305-4470/24/i=4/a=003} {\bibfield
  {journal} {\bibinfo  {journal} {J. Phys. A}\ }\textbf {\bibinfo {volume}
  {24}},\ \bibinfo {pages} {L175} (\bibinfo {year} {1991})}\BibitemShut
  {NoStop}%
\bibitem [{\citenamefont {Kernaghan}\ and\ \citenamefont
  {Peres}(1995)}]{Kernaghan19951}%
  \BibitemOpen
  \bibfield  {author} {\bibinfo {author} {\bibfnamefont {M.}~\bibnamefont
  {Kernaghan}}\ and\ \bibinfo {author} {\bibfnamefont {A.}~\bibnamefont
  {Peres}},\ }\href
  {http://www.sciencedirect.com/science/article/pii/037596019500012R}
  {\bibfield  {journal} {\bibinfo  {journal} {Phys. Lett. A}\ }\textbf
  {\bibinfo {volume} {198}},\ \bibinfo {pages} {1 } (\bibinfo {year}
  {1995})}\BibitemShut {NoStop}%
\bibitem [{\citenamefont {Cabello}\ \emph {et~al.}(1996)\citenamefont
  {Cabello}, \citenamefont {Estebaranz},\ and\ \citenamefont
  {Garc\'ia-Alcaine}}]{Cabello1996183}%
  \BibitemOpen
  \bibfield  {author} {\bibinfo {author} {\bibfnamefont {A.}~\bibnamefont
  {Cabello}}, \bibinfo {author} {\bibfnamefont {J.~M.}\ \bibnamefont
  {Estebaranz}}, \ and\ \bibinfo {author} {\bibfnamefont {G.}~\bibnamefont
  {Garc\'ia-Alcaine}},\ }\href
  {http://www.sciencedirect.com/science/article/pii/037596019600134X}
  {\bibfield  {journal} {\bibinfo  {journal} {Phys. Lett. A}\ }\textbf
  {\bibinfo {volume} {212}},\ \bibinfo {pages} {183 } (\bibinfo {year}
  {1996})}\BibitemShut {NoStop}%
\bibitem [{\citenamefont {Yu}\ and\ \citenamefont
  {Oh}(2012)}]{PhysRevLett.108.030402}%
  \BibitemOpen
  \bibfield  {author} {\bibinfo {author} {\bibfnamefont {S.}~\bibnamefont
  {Yu}}\ and\ \bibinfo {author} {\bibfnamefont {C.~H.}\ \bibnamefont {Oh}},\
  }\href {http://link.aps.org/doi/10.1103/PhysRevLett.108.030402} {\bibfield
  {journal} {\bibinfo  {journal} {Phys. Rev. Lett.}\ }\textbf {\bibinfo
  {volume} {108}},\ \bibinfo {pages} {030402} (\bibinfo {year}
  {2012})}\BibitemShut {NoStop}%
\bibitem [{\citenamefont {Waegell}\ and\ \citenamefont
  {Aravind}(2013)}]{PhysRevA.88.012102}%
  \BibitemOpen
  \bibfield  {author} {\bibinfo {author} {\bibfnamefont {M.}~\bibnamefont
  {Waegell}}\ and\ \bibinfo {author} {\bibfnamefont {P.~K.}\ \bibnamefont
  {Aravind}},\ }\href {http://link.aps.org/doi/10.1103/PhysRevA.88.012102}
  {\bibfield  {journal} {\bibinfo  {journal} {Phys. Rev. A}\ }\textbf {\bibinfo
  {volume} {88}},\ \bibinfo {pages} {012102} (\bibinfo {year}
  {2013})}\BibitemShut {NoStop}%
\bibitem [{\citenamefont {Pavi\v{c}i\'c}\ \emph {et~al.}(2010)\citenamefont
  {Pavi\v{c}i\'c}, \citenamefont {Megill},\ and\ \citenamefont
  {Merlet}}]{Pavicic20102122}%
  \BibitemOpen
  \bibfield  {author} {\bibinfo {author} {\bibfnamefont {M.}~\bibnamefont
  {Pavi\v{c}i\'c}}, \bibinfo {author} {\bibfnamefont {N.~D.}\ \bibnamefont
  {Megill}}, \ and\ \bibinfo {author} {\bibfnamefont {J.-P.}\ \bibnamefont
  {Merlet}},\ }\href
  {http://www.sciencedirect.com/science/article/pii/S0375960110002999}
  {\bibfield  {journal} {\bibinfo  {journal} {Phys. Lett. A}\ }\textbf
  {\bibinfo {volume} {374}},\ \bibinfo {pages} {2122 } (\bibinfo {year}
  {2010})}\BibitemShut {NoStop}%
\bibitem [{\citenamefont {Toh}\ and\ \citenamefont
  {Zainuddin}(2010)}]{Toh20104834}%
  \BibitemOpen
  \bibfield  {author} {\bibinfo {author} {\bibfnamefont {S.}~\bibnamefont
  {Toh}}\ and\ \bibinfo {author} {\bibfnamefont {H.}~\bibnamefont
  {Zainuddin}},\ }\href
  {http://www.sciencedirect.com/science/article/pii/S0375960110013587}
  {\bibfield  {journal} {\bibinfo  {journal} {Phys. Lett. A}\ }\textbf
  {\bibinfo {volume} {374}},\ \bibinfo {pages} {4834 } (\bibinfo {year}
  {2010})}\BibitemShut {NoStop}%
\bibitem [{\citenamefont {Cabello}(2003)}]{PhysRevLett.90.190401}%
  \BibitemOpen
  \bibfield  {author} {\bibinfo {author} {\bibfnamefont {A.}~\bibnamefont
  {Cabello}},\ }\href {http://link.aps.org/doi/10.1103/PhysRevLett.90.190401}
  {\bibfield  {journal} {\bibinfo  {journal} {Phys. Rev. Lett.}\ }\textbf
  {\bibinfo {volume} {90}},\ \bibinfo {pages} {190401} (\bibinfo {year}
  {2003})}\BibitemShut {NoStop}%
\bibitem [{\citenamefont {Aravind}(2003)}]{PhysRevA.68.052104}%
  \BibitemOpen
  \bibfield  {author} {\bibinfo {author} {\bibfnamefont {P.~K.}\ \bibnamefont
  {Aravind}},\ }\href {http://link.aps.org/doi/10.1103/PhysRevA.68.052104}
  {\bibfield  {journal} {\bibinfo  {journal} {Phys. Rev. A}\ }\textbf {\bibinfo
  {volume} {68}},\ \bibinfo {pages} {052104} (\bibinfo {year}
  {2003})}\BibitemShut {NoStop}%
\bibitem [{\citenamefont {Stapp}(1971)}]{PhysRevD.3.1303}%
  \BibitemOpen
  \bibfield  {author} {\bibinfo {author} {\bibfnamefont {H.~P.}\ \bibnamefont
  {Stapp}},\ }\href {\doibase 10.1103/PhysRevD.3.1303} {\bibfield  {journal}
  {\bibinfo  {journal} {Phys. Rev. D}\ }\textbf {\bibinfo {volume} {3}},\
  \bibinfo {pages} {1303} (\bibinfo {year} {1971})}\BibitemShut {NoStop}%
\bibitem [{\citenamefont {Eberhard}(1977)}]{Eberhard1977}%
  \BibitemOpen
  \bibfield  {author} {\bibinfo {author} {\bibfnamefont {P.}~\bibnamefont
  {Eberhard}},\ }\href {\doibase 10.1007/BF02726212} {\bibfield  {journal}
  {\bibinfo  {journal} {Il Nuov. Cim. B}\ }\textbf {\bibinfo {volume} {38}},\
  \bibinfo {pages} {75} (\bibinfo {year} {1977})}\BibitemShut {NoStop}%
\bibitem [{\citenamefont {Abbott}\ \emph {et~al.}(2014)\citenamefont {Abbott},
  \citenamefont {Calude},\ and\ \citenamefont {Svozil}}]{PhysRevA.89.032109}%
  \BibitemOpen
  \bibfield  {author} {\bibinfo {author} {\bibfnamefont {A.~A.}\ \bibnamefont
  {Abbott}}, \bibinfo {author} {\bibfnamefont {C.~S.}\ \bibnamefont {Calude}},
  \ and\ \bibinfo {author} {\bibfnamefont {K.}~\bibnamefont {Svozil}},\ }\href
  {\doibase 10.1103/PhysRevA.89.032109} {\bibfield  {journal} {\bibinfo
  {journal} {Phys. Rev. A}\ }\textbf {\bibinfo {volume} {89}},\ \bibinfo
  {pages} {032109} (\bibinfo {year} {2014})}\BibitemShut {NoStop}%
\bibitem [{\citenamefont {Spekkens}(2005)}]{PhysRevA.71.052108}%
  \BibitemOpen
  \bibfield  {author} {\bibinfo {author} {\bibfnamefont {R.~W.}\ \bibnamefont
  {Spekkens}},\ }\href {\doibase 10.1103/PhysRevA.71.052108} {\bibfield
  {journal} {\bibinfo  {journal} {Phys. Rev. A}\ }\textbf {\bibinfo {volume}
  {71}},\ \bibinfo {pages} {052108} (\bibinfo {year} {2005})}\BibitemShut
  {NoStop}%
\bibitem [{\citenamefont {Clauser}\ \emph {et~al.}(1969)\citenamefont
  {Clauser}, \citenamefont {Horne}, \citenamefont {Shimony},\ and\
  \citenamefont {Holt}}]{PhysRevLett.23.880}%
  \BibitemOpen
  \bibfield  {author} {\bibinfo {author} {\bibfnamefont {J.~F.}\ \bibnamefont
  {Clauser}}, \bibinfo {author} {\bibfnamefont {M.~A.}\ \bibnamefont {Horne}},
  \bibinfo {author} {\bibfnamefont {A.}~\bibnamefont {Shimony}}, \ and\
  \bibinfo {author} {\bibfnamefont {R.~A.}\ \bibnamefont {Holt}},\ }\href
  {\doibase 10.1103/PhysRevLett.23.880} {\bibfield  {journal} {\bibinfo
  {journal} {Phys. Rev. Lett.}\ }\textbf {\bibinfo {volume} {23}},\ \bibinfo
  {pages} {880} (\bibinfo {year} {1969})}\BibitemShut {NoStop}%
\bibitem [{\citenamefont {Fine}(1982)}]{PhysRevLett.48.291}%
  \BibitemOpen
  \bibfield  {author} {\bibinfo {author} {\bibfnamefont {A.}~\bibnamefont
  {Fine}},\ }\href {\doibase 10.1103/PhysRevLett.48.291} {\bibfield  {journal}
  {\bibinfo  {journal} {Phys. Rev. Lett.}\ }\textbf {\bibinfo {volume} {48}},\
  \bibinfo {pages} {291} (\bibinfo {year} {1982})}\BibitemShut {NoStop}%
\bibitem [{\citenamefont {Beltrametti}\ and\ \citenamefont
  {Cassinelli}(1981)}]{Beltrametti:QM81}%
  \BibitemOpen
  \bibfield  {author} {\bibinfo {author} {\bibfnamefont {E.}~\bibnamefont
  {Beltrametti}}\ and\ \bibinfo {author} {\bibfnamefont {G.}~\bibnamefont
  {Cassinelli}},\ }\href@noop {} {\emph {\bibinfo {title} {The Logic of Quantum
  Mechanics}}}\ (\bibinfo  {publisher} {Addison -Wesley},\ \bibinfo {year}
  {1981})\BibitemShut {NoStop}%
\end{thebibliography}%

\end{document}